\documentclass[aps,prd,amsmath,amssymb,preprintnumbers,superscriptaddress,a4paper,11pt]{revtex4-1}
\pdfoutput=1
\usepackage{amsthm}
\usepackage{graphicx}
\usepackage[ margin=5pt, font=small,labelfont=bf, justification=raggedright]{caption}
\usepackage{url}
\usepackage[bookmarks, pagebackref=false]{hyperref}
\usepackage{color}
	\definecolor{rossoCP3}{cmyk}{0,.88,.77,.40}
		\definecolor{graa}{rgb}{0.8,0.8,0.8}
		\definecolor{blaa}{rgb}{0.2,0.2,0.6}
		\hypersetup{
			colorlinks,
			bookmarksopen,
			bookmarksnumbered,
			citecolor=blaa, 		
			linkcolor=rossoCP3,	
			urlcolor=rossoCP3,			
			}
\usepackage{dcolumn}
\usepackage{bm}
\usepackage{bbm}
\usepackage{pxfonts}

\usepackage{amsmath}
\usepackage{amsfonts}

\usepackage{dcolumn}
\usepackage{bm}
\usepackage{amssymb}
\usepackage{latexsym}
\usepackage{color}

\usepackage{epsfig}
\usepackage{placeins}

\usepackage{youngtab}
\usepackage{slashed}
\usepackage{subfig}

\newcommand{\be}{\begin{equation}}
\newcommand{\ee}{\end{equation}}
\newcommand{\bea}{\begin{eqnarray}}
\newcommand{\eea}{\end{eqnarray}}
\newcommand{\nn}{\nonumber}

\begin{document}

\title{\Large  125 GeV Higgs from a chiral-techniquark model}
\author{Stefano Di Chiara}
\email{stefano.dichiara@helsinki.fi}
\affiliation{Department of Physics \& Helsinki Institute of Physics, P.O. Box 64,
FI-000140, University of Helsinki, Finland}
\author{Roshan Foadi}
\email{roshan.foadi@gmail.com}
\affiliation{Department of Physics, University of Jyv\"askyl\"a, P.O. Box 35, 
FI-40014, University of Jyv\"askyl\"a, Finland}
\affiliation{Department of Physics \& Helsinki Institute of Physics, P.O. Box 64,
FI-000140, University of Helsinki, Finland}
\author{Kimmo Tuominen}
\email{kimmo.i.tuominen@helsinki.fi}
\affiliation{Department of Physics \& Helsinki Institute of Physics, P.O. Box 64,
FI-000140, University of Helsinki, Finland}

\begin{abstract}
We consider the spin-zero spectrum of a strongly coupled gauge theory. In particular, we focus on the dynamical mass of the isosinglet scalar resonance in the presence of a four-fermion interaction external to the gauge dynamics. This is motivated by the extended technicolor framework for dynamical electroweak symmetry breaking. Applying the large-$N$ limit, we sum all the leading-order contributions, and find that the corrections to the mass of the isosinglet scalar resonance can be large, potentially reducing its value from ${\cal O}(1)$ TeV to the observed value of 125 GeV. 
\end{abstract}

\maketitle

\section{Introduction}

In light of current collider data \cite{Aad:2012tfa, Chatrchyan:2012ufa}, the Standard Model (SM) of elementary particle interactions remains as an accurate
description of nature \cite{Azatov:2012bz, Giardino:2013bma, Alanne:2013dra}. Nevertheless, observational evidence for the dark matter abundance, lack of a mechanism for generating the matter-antimatter asymmetry, and no understanding of the dynamics underlying the flavor patterns, continue to provide impetus for development and analysis of beyond-the-Standard-Model scenarios. The need for such scenarios is further augmented by the hierarchy and naturality problems associated with a fundamental scalar field \cite{'tHooft:1979bh, Farina:2013mla, Heikinheimo:2013fta,Craig:2013xia,Antipin:2013exa}. 

A traditional model building paradigm to address the naturality problem is to replace the scalar sector with new fermions charged under
new gauge dynamics \cite{Weinberg:1975gm, Susskind:1978ms}. These technicolor (TC), or composite Higgs, models also provide dark matter 
candidates, and affect the electroweak phase transition to make electroweak baryogensis possible. Extended technicolor 
models (ETC)\cite{ Dimopoulos:1979es, Eichten:1979ah}  provide ways to understand generational hierarchies \cite{Appelquist:2003hn}. However, TC model building is challenging due to our limited 
understanding of strong dynamics from first principles. Consequently, a large amount of common wisdom concerning the 
phenomenological details of dynamical electroweak symmetry breaking rests on the naive scaling arguments from what we 
know about QCD. The most profound example of this is the scalar spectrum: if we consider the sigma meson, i.e. the 
resonance denoted $f_0(500)$ by the particle data group \cite{Beringer:1900zz}, and scale from QCD with $f_\pi\simeq 100$ MeV to a 
technicolor theory with $F_\pi\simeq 250$ GeV, we find that the corresponding scalar has mass $M_\sigma\simeq 1250$ GeV. 
Small refinements arise from taking into account also possibly different number of colours and different fermion 
representations. But still one is typically bound to the conclusion that the scalar mass is in the TeV range, thus much heavier than the recently observed scalar boson at 125 GeV  \cite{Aad:2012tfa, Chatrchyan:2012ufa}. Refinements to naive scaling from QCD arise in theories which are nearly conformal, also called walking tehcnicolor theories. Various arguments for the existence of a light scalar in these theories have been provided in the literature \cite{Yamawaki:1985zg,Bando:1986bg,Miransky:1989nu, Dietrich:2005jn,Chacko:2012sy,Elander:2010wd,Alho:2013dka}. 

Whichever is the case, one should be careful when comparing the estimates of the dynamical mass directly with the mass 
of the Higgs boson observed at the LHC experiments. An element which is often overlooked is that the scaling described 
above applies to strongly interacting theories in isolation. When coupled with the full electroweak sector, there can be 
important contributions affecting the conclusions \cite{Foadi:2012bb}. Here again, QCD provides a guide to orient our thinking: the mass 
difference between charged and neutral pions arises from the electromagnetic interaction, and is explicitly and rigorously 
given in the chiral limit by 
\bea
\Delta m_\pi^2 &\equiv& m^2_{\pi^+}-m_{\pi0}^2 \nn \\
&=& -\frac{3\alpha_{\rm{EM}}}{4\pi f_\pi^2}\int_0^\infty ds\, s\, \ln s (\rho_V(s)-\rho_A(s)),
\eea
where $\rho_{V,A}(s)$ are the vector and axial vector spectral functions. Experimentally, this mass difference is determined 
to be (35 MeV)$^2$ for QCD pions. The relative smallness, $\Delta m_\pi^2/m_\pi^2\simeq 0.06$, is entirely due to the 
weakness of the electroweak interaction. If instead we had $\alpha_{\rm{EM}}\sim 1/(4\pi)$, the relative magnitude 
would be 0.7, i.e. of ${\cal O}(1)$ relative to the dynamical mass from strong interactions. Here the key element is that in the case of QCD the electromagnetic interaction is completely external to the 
strong dynamics, while in the case of technicolor it is the electroweak theory itself that plays the role of an external interaction. For the 
electroweak sector the couplings, especially the Yukawa coupling of the top quark, are larger than the electromagnetic coupling.

In the ETC framework the coupling between the technicolor sector and SM matter fields is modelled at 
low energies via four-fermion operators. In this paper we consider the lightest isosinglet scalar boson arising from a strongly interacting 
sector, and compute the effect on its mass from the four-fermion operator responsible for the top mass. We find that the corrections can be 
large and potentially reduce the scalar resonance mass from ${\cal O}(1)$ TeV to $\sim$125 GeV. We work with a simple setup, consisting of a
chiral model for the TC sector, and a single four-fermion coupling. This is sufficient for exhibiting how the correction to the 
dynamical mass arises and what its expected magnitude is. To account for the fermion mass generation as well as to 
address the precision electroweak data, a more detailed sector of four-fermion interactions would be needed. We leave such 
investigation for future work and concentrate here only on the determination of the Higgs mass in this framework, which is 
currently the most important feature in light of current and future LHC data.

The paper is structured as follows: In section~\ref{Sec:model} we introduce the model and in section~\ref{Sec:EWmatch} we 
discuss the matching with the electroweak theory, the top mass and the Fermi coupling constant. The main result, i.e. the analysis of the mass of the scalar 
isosinglet, is presented in section~\ref{Sec:scalarmass}. In section~\ref{Sec:ST} we briefly discuss the oblique electroweak parameters and the implications of our analysis on the underlying strong dynamics. In section~\ref{Sec:conclusions} we present our conclusions and outlook 
for future work.

\section{Chiral techniquark model}
\label{Sec:model}
We consider TC theories in which the lightest resonances are composed of fermions belonging to 
one weak technidoublet $Q\equiv (U,D)$. This does not imply that there could not exist additional technifermions, 
as these can be heavy enough to be decoupled from the lightest resonances. In QCD, for instance, the sigma meson contains the $u$ and $d$ quarks but not the $s$ quark, which is heavier and decoupled. 
To discuss phenomenology at the energies explored by the LHC, we consider the effective Lagrangian
\begin{equation}
{\cal L}={\cal L}_{\overline {\rm SM}} + {\cal L}_{\rm TC} + {\cal L}_{\rm ETC}\ ,
\label{Eq:lagfull}
\end{equation}
where ${\cal L}_{\overline {\rm SM}} $ is the SM Lagrangian without Higgs and Yukawa terms, ${\cal L}_{\rm TC}$ is a model Lagrangian accounting for the non-perturbative TC dynamics, and ${\cal L}_{\rm ETC}$ contains interactions mediated by exchanges of heavy ETC gauge bosons, which we assume to exceed the energy reach of the LHC. The full Lagrangian is invariant under the SM gauge group, and the electroweak symmetry is spontaneously broken by the TC force to electromagnetism.

The ${\cal L}_{\rm TC}$ part of the full Lagrangian is given by a symmetry-breaking chiral-techniquark model, containing both composite resonances and the techniquarks. Among the resonances, we assume the lightest ones to be the massless technipion isospin triplet $\Pi^a$,  which becomes the longitudinal component of the $W$ and $Z$ bosons, and the scalar singlet $H$. Our main goal is to determine whether the latter can be a candidate for the recently observed 125 GeV resonance. In order to account for compositeness, we take the kinetic terms for $\Pi^a$ and $H$ to be radiatively generated. Explicitly,
\begin{equation}
{\cal L}_{\rm TC} = \overline{Q}_L\ i\slashed{D} Q_L + \overline{U}_R\ i\slashed{D} U_R + \overline{D}_R\ i\slashed{D} D_R  
-M\left(1+\frac{y}{v}H+\cdots\right)\left(\overline{Q}_L\Sigma Q_R+\overline{Q}_R\Sigma^\dagger Q_L\right) -\frac{m^2}{2} H^2 +\cdots\ ,
\label{eq:TC}
\end{equation}
where $\Sigma\equiv\exp\left(i\Pi^a\tau^a/v\right)$, and $\tau^a$, $a=1,2,3$, are the Pauli matrices. 
The dynamical constituent mass of the techniquarks is denoted by $M$. The ellipses denote higher-order interactions, which we assume to give negligible contribution to the masses and decay constants. 
Note that we have suppressed the TC gauge index in the techniquarks, and that the covariant derivatives are with respect to the SM gauge interactions. 

In the ETC sector we only consider the four-fermion operator which allows the top quark to interact with the techniquark condensate and acquire mass,
\begin{equation}
{\cal L}_{\rm ETC} = 2 G\left(\overline{q}_L^i t_R \overline{U}_R Q_{iL} + \overline{Q}_L^i U_R \overline{t}_R q_{iL}\right) \ ,
\label{eq:ETC}
\end{equation}
where $q\equiv (t,b)$ is the top-bottom doublet, and $i$ is a weak isospin index. The inclusion of this single interaction term
is sufficient for the study of the generic effect of four-fermion interactions on the scalar mass.
A more complete discussion of fermion mass patterns and precision electroweak observables would require additional operators. We do not consider such refinements in this paper.
\subsection{Cutoff and confinement}
Using the Lagrangian of (\ref{Eq:lagfull}), we can compute observables such as the mass of the scalar singlet $H$ and the technipion decay constant. We do so in the large-$N$ and large-$N_c$ limit, with $N/N_c$ finite: here $N$ is the dimension of the techniquark representation under TC, whereas $N_c=3$ is the dimension of the quark representation under QCD. We assume that the loop integrals are finite, and that their absolute size has a physical meaning. Practically, this means that we use a physical cutoff. However, there are two relevant mass scales, the mass of the lightest ETC gauge boson, ${M_{\rm ETC}}$, and the scale of compositeness, ${\Lambda_{\rm TC}}$, involved. It is not clear which of these should be used as a cutoff. A reasonable approach would consist in using the smaller mass scale, which is expected to be ${\Lambda_{\rm TC}}$. This, however, implies losing information from the dynamics occurring between ${\Lambda_{\rm TC}}$ and ${M_{\rm ETC}}$. Furthermore, it is well know that making the techniquark loop integrals finite with a sharp cutoff does not account for confinement, as the fermion propagators go on-shell for sufficiently large external momenta.  A solution to both problems is provided by models of confinement \cite{Efimov:1993zg}. To briefly review these, let $S(x_1-x_2)$ be the Green function of a confined techniquark in the vacuum technigluon field. This can be represented as
\begin{equation}
S(x_1-x_2)=-\int\frac{d^4 p}{(2\pi)^4}\, i\, \rho(\slashed{p})\, e^{-i\, p\cdot (x_1-x_2)}\ .
\label{Eq:Confinement}
\end{equation} 
Confinement means that the techniquark field has no asymptotically free states which describe non-interacting free particles. This, in turn, implies that the function $\rho(\slashed{p})$ is everywhere holomorphic. Therefore, we can use the Cauchy representation
\begin{equation}
\rho(\slashed{p}) = -\int_L \frac{dM}{2\pi i}\, \frac{\rho(M)}{\slashed{p}-M}\ ,
\label{Eq:Cauchy}
\end{equation}
where $L$ is a closed contour around $\slashed{p}=0$. Inserting (\ref{Eq:Cauchy}) in (\ref{Eq:Confinement}) gives
\begin{equation}
S(x_1-x_2)=\int_L \frac{dM}{2\pi i}\, \rho(M)\, S(x_1-x_2,M)\ ,
\label{Eq:S}
\end{equation}
where $S(x_1-x_1,M)$ is the propagator of a free techniquark of mass $M$. The techniquark condensate is
\begin{equation}
\langle \overline{Q}Q \rangle = -\, {\rm Tr}\, S(0) = \int_L \frac{dM}{2\pi i}\, \rho(M)\,  \langle \overline{Q}Q \rangle_M\ ,
\end{equation}
where $\langle \overline{Q}Q \rangle_M\equiv -\, {\rm Tr}\, S(0,M)$ is the condensate of the free techniquark field. This means that $\rho(M)$ can be interpreted as a distribution density of techniquark masses in the vacuum technigluon field. We can generalize (\ref{Eq:S}) to Green functions of $n$ external technihadrons,
\begin{equation}
{\cal T}_n(x_1,x_2,\dots,n)\, =\, 
\int_L \frac{dM}{2\pi i}\, \rho(M)\, {\rm Tr}\, \Big( \Gamma_1\, S(x_1-x_2,M)\cdots \Gamma_n\, S(x_n-x_1,M) \Big)
\end{equation}
where $\Gamma_k$ are matrices in Dirac space. In each propagator the techniquark constituent mass is "smeared" by the distribution density $\rho(M)$, which playes a double role: First, it provides convergence of the integrals in momentum space, within a domain the size of which is
determined by the
mass scale $\Lambda_{\rm TC}$. Second, it prevents the techniquark propagators to go on-shell, thus removing unwanted production of free techniquarks. The function $\rho(M)$ cannot be computed exactly, 
and an ansatz must be made based on first principles and phenomenological constraints.

If we use a distribution density $\rho(M)$ to smear the integrals over techniquarks, we may cutoff the ${\cal L}_{\rm TC} + {\cal L}_{\rm ETC}$ theory at ${M_{\rm ETC}}$.  The integrals over SM quarks are cutoff at ${M_{\rm ETC}}$, whereas the integrals over techniquarks are naturally finite and of the order of ${\Lambda_{\rm TC}}$. No techniquark propagator can go on-shell, and confinement is therefore guaranteed. Clearly we must choose an appropriate function $\rho(M)$, and integrals are unavoidably more difficult to evaluate than the standard loop integrals. However in our analysis we are only interested in small external momenta, and no internal techniquark propagator can go on-shell, even without a distribution density. 
If a distribution density was used, the consequence would be that is that it makes the integrals finite and of the order of ${\Lambda_{\rm TC}}$. Therefore, we make the approximation of using a sharp cutoff 
${\Lambda_{\rm TC}}$ for the loop integrals over techniquark momenta, rather than a distribution density, while still cutting off the SM-fermion loop integrals at ${M_{\rm ETC}}$. This approach makes the dynamics between ${\Lambda_{\rm TC}}$ and ${M_{\rm ETC}}$ contribute to the low-energy observables, and, as we shall see, preserves the global symmetries of the Lagrangian. Of course, working with a cutoff carries some ambiguity when the external momentum $q^2$  is non-zero. 
However, for the observables we consider in our analysis, $q^2$ is either exactly zero or very small in comparison to the techniquark mass squared. The standard loop integrals used in our computations are explicitly given in the appendix.
\section{Matching with the electroweak theory} \label{Sec:EWmatch}
\begin{figure}[htb]
\includegraphics[width=4.5in]{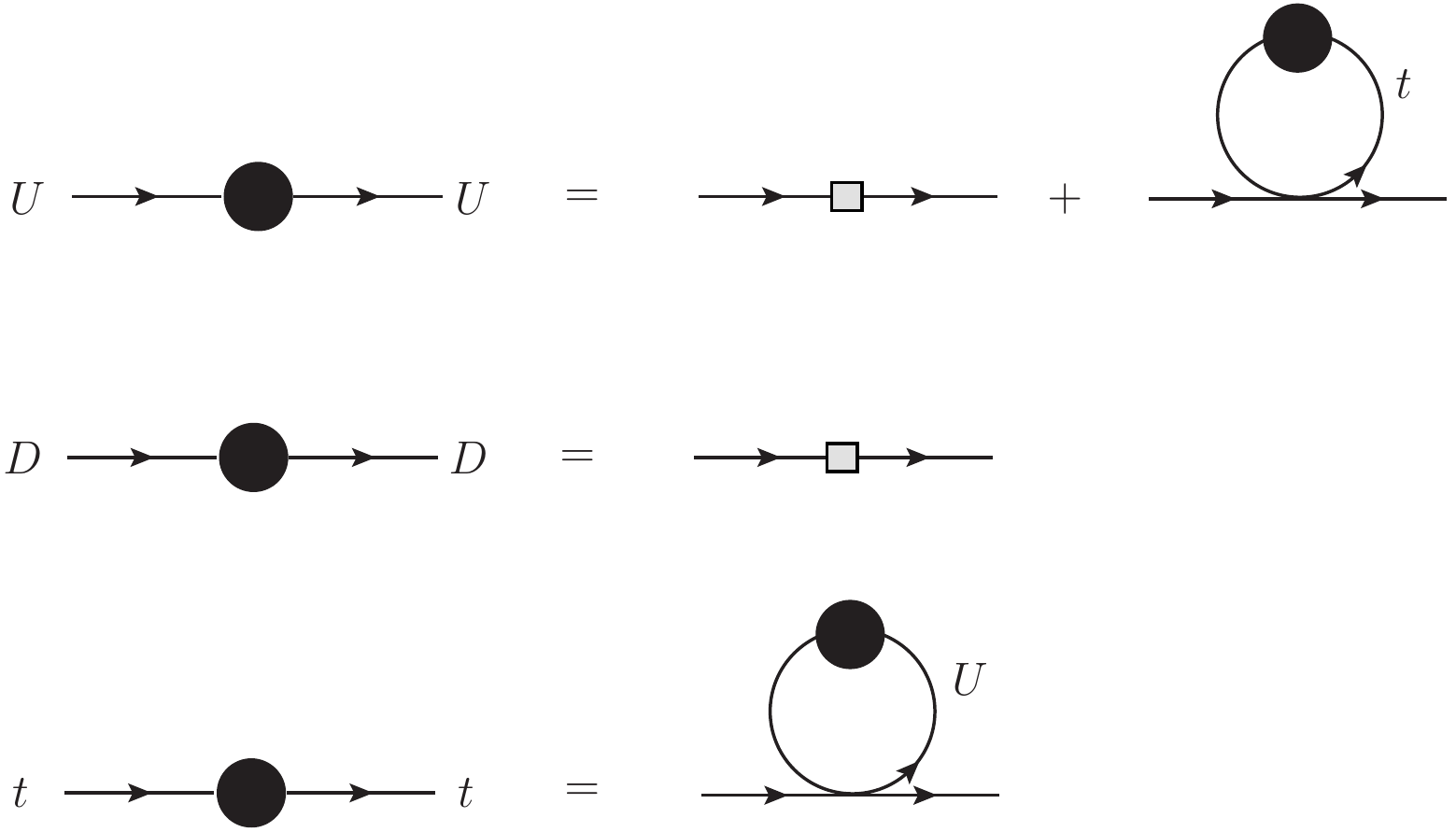}
\caption{Diagrams contributing to the fermion masses at leading order in $N$ and $N_c$. $M_D$ is entirely given by the tree-level term, whereas $M_U$ receives both tree-level and loop contributions. Note that the bottom quark is massless in our approximation.}
\label{Fig:Gap}
\end{figure}
In our approximation the top quark acquires mass, whereas the bottom quark is massless. In the large-$N$, $N_c$ limit, with $N/N_c$ finite, the fermion masses are given by the diagrams of Fig.~\ref{Fig:Gap}. These correspond to the coupled gap equations
\begin{eqnarray}
&& M_U = M + 4\, N_c\, G\, M_t\, I_1^{M_{\rm ETC}}(M_t)\ , \nonumber \\
&& M_t = 4\, N\, G\, M_U\, I_1^{\Lambda_{\rm TC}}(M_U)\ ,
\label{eq:Gap}
\end{eqnarray}
and $M_D=M$, where $I_1^\Lambda(M)$ is defined in (\ref{Eq:I1}). As motivated in the last section, here and in the computations below, the integrals over techniquarks are cutoff at 
${\Lambda_{\rm TC}}$, whereas the integrals over SM fermions are cutoff at the ETC mass scale ${M_{\rm ETC}}$. We define a measure of isospin mass splitting in the techniquark sector as
\begin{equation}
\delta\equiv \frac{M_U-M}{M}\ .
\label{Eq:Delta}
\end{equation}
The gap equations can be solved numerically for $G$ and $\delta$.

\begin{figure}[!t]
\includegraphics[width=6.2in]{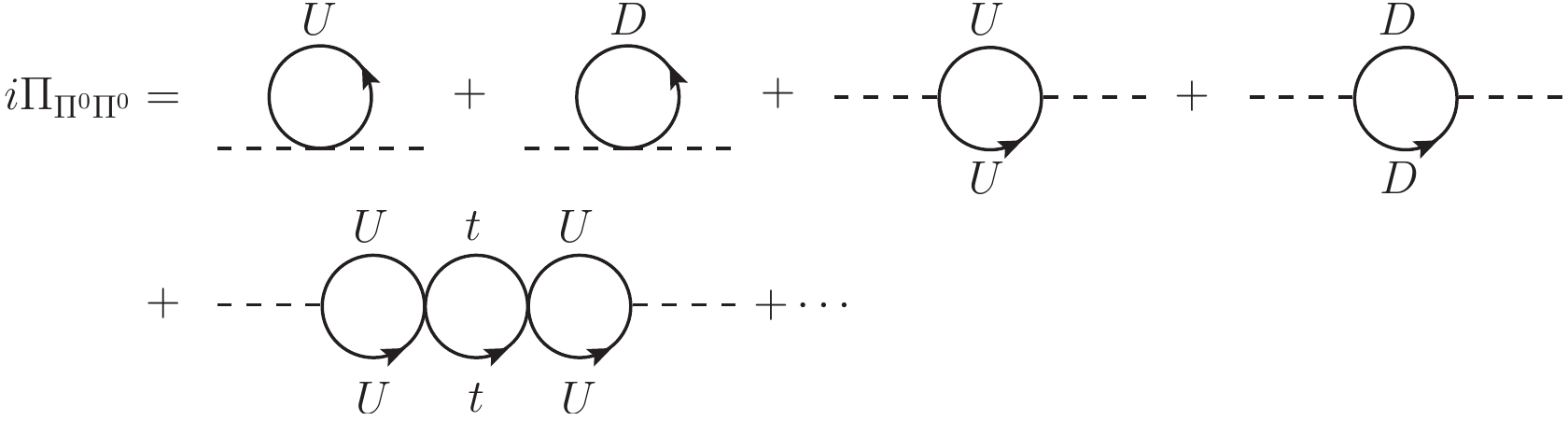}
\caption{Diagrams contributing to the neutral technipion self-energy at leading order in $N$ and $N_c$.}
\label{Fig:NeutralPi}
\end{figure}

The charged and neutral technipion are the Goldstone bosons "eaten" by the longitudinal components of the $W$ and $Z$ boson, respectively. To ensure that the cutoff scheme respects chiral symmetry, we need to check that these absorbed Goldstone bosons are massless.
Let us start with the neutral technipion. Its self-energy $\Sigma_{\Pi^0\Pi^0}$ is given by the infinite sum of diagrams shown in Fig.~\ref{Fig:NeutralPi}. 

This leads to the expression
\begin{equation}
\Sigma_{\Pi^0\Pi^0} = -\frac{4\, N\, M}{v^2}\left(M_U\, I_1^{\Lambda_{\rm TC}}(M_U)+M\, \ I_1^{\Lambda_{\rm TC}}(M)\right)
+\frac{N\, M^2}{v^2}\left(\chi_D + \chi_U\right)
+\frac{N\, M^2}{v^2}\frac{N\, N_c\, G^2\, \chi_U^2\, \chi_t}{1-N\, N_c\, G^2\, \chi_U\, \chi_t}\ ,
\end{equation}
where
\begin{equation}
i\ \chi_X\equiv -\int\frac{d^4k}{(2\pi)^4} {\rm Tr}\ \gamma_5 \frac{i\Big(\slashed{k}+M_X\Big)}{k^2-M_X^2}\gamma_5 \frac{i\Big(\slashed{k}+\slashed{q}+M_X\Big)}{(k+q)^2-M_X^2}\ .
\end{equation}
Assuming $q^2\ll M_X^2$ the integral is given by

\begin{eqnarray}
&& \chi_X = 2\, q^2\, I_2^{\Lambda_{\rm TC}}(M_X)+4\, I_1^{\Lambda_{\rm TC}}(M_X)\ ,\ \ X=U,D \\
&& \chi_t = 2\, q^2\, I_2^{M_{\rm ETC}}(M_t) + 4\, I_1^{M_{\rm ETC}}(M_t)\ ,
\end{eqnarray}
where $I^\Lambda(M)$ is defined in (\ref{Eq:I2}).
Using these expressions, $\Sigma_{\Pi^0\Pi^0}$ becomes
\begin{equation}
\Sigma_{\Pi^0\Pi^0} = \frac{ N\, M^2}{v^2}\left(2\, I_2^{\Lambda_{\rm TC}}(M) q^2
-4 \frac{M_U}{M}\ I_1^{\Lambda_{\rm TC}}(M_U)
+\frac{\chi_U}{1-N\, N_c\, G^2\, \chi_U\, \chi_t}\right)\ .
\end{equation}
Using the gap equations it is rather straightforward to show that this expression vanishes at $q^2=0$, 
proving that the neutral technipion is massless.

\begin{figure}[!t]
\includegraphics[width=6.7in]{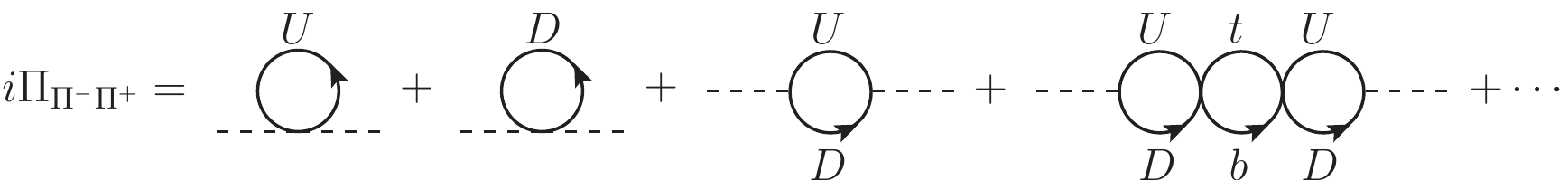}
\caption{Diagrams contributing to the charged technipion self-energy at leading order in $N$ and $N_c$.}
\label{Fig:ChargedPi}
\end{figure}
Then let us turn to the charged technipion. The computation showing that it is massless is similar to the one for the neutral technipion. The charged-technipion self-energy is given by the diagrams of Fig.~\ref{Fig:ChargedPi}, and summing the infinite series gives
\begin{equation}
\Sigma_{\Pi^-\Pi^+} = -\frac{4\, N\, M}{v^2}\left(M_U\, I_1^{\Lambda_{\rm TC}}(M_U)+M\, \ I_1^{\Lambda_{\rm TC}}(M)\right)
+\frac{2\, N\, M^2}{v^2}\left(\chi_{UD}
+\frac{N\, N_c\, G^2\, \chi_{UD}^2\, \zeta_{tb}}{1-4 N\, N_c\, G^2\, \zeta_{UD}\, \zeta_{tb}}\right)\ ,
\end{equation}
where
\begin{eqnarray}
&& i\ \chi_{UD}\equiv -\int\frac{d^4k}{(2\pi)^4} {\rm Tr}\ \gamma_5 \frac{i\Big(\slashed{k}+M\Big)}{k^2-M^2}\gamma_5 \frac{i\Big(\slashed{k}+\slashed{q}+M_U\Big)}{(k+q)^2-M_U^2}\ , \nonumber \\
&& i\ \zeta_{XY}\equiv \int\frac{d^4k}{(2\pi)^4} {\rm Tr}\ P_L \frac{i\Big(\slashed{k}+M_Y\Big)}{k^2-M_Y^2} P_R \frac{i\Big(\slashed{k}+\slashed{q}+M_X\Big)}{(k+q)^2-M_X^2}\ .
\end{eqnarray}
Evaluating the integrals gives
\begin{eqnarray}
&& \chi_{UD} = 2\left(q^2-\delta^2 M^2\right) I_2^{\Lambda_{\rm TC}}(M_U,M)+2 \left(I_1^{\Lambda_{\rm TC}}(M_U)+I_1^{\Lambda_{\rm TC}}(M)\right)\ ,
\end{eqnarray}
and
\begin{eqnarray}
&& \zeta_{UD} = \left(q^2-M_U^2-M^2\right) I_2^{\Lambda_{\rm TC}}(M_U,M)+I_1^{\Lambda_{\rm TC}}(M_U)+I_1^{\Lambda_{\rm TC}}(M)\ ,\nonumber \\
&& \zeta_{tb} = \left(q^2-M_t^2\right) I_2^{M_{\rm ETC}}(M_t,0)+ I_1^{M_{\rm ETC}}(M_t)+I_1^{M_{\rm ETC}}(0)\ .
\end{eqnarray}
The function $I_2^\Lambda(M_1,M_2)$ is defined in Eq. (\ref{Eq:I2b}).
Using these expressions, $\Sigma_{\Pi^-\Pi^+}$ at zero momentum becomes
\begin{equation}
\Sigma_{\Pi^+\Pi^-}\left(q^2=0\right) = \left.
\frac{ N\, M^2}{2 v^2}\frac{ \chi_{UD}}{\zeta_{UD}}\left(
-4 \frac{M_U}{M}\ I_1^{\Lambda_{\rm TC}}(M_U)
+\frac{\chi_{UD}}{1-4\,N\, N_c\, G^2\, \zeta_{UD}\, \zeta_{tb}}\right)\right\vert_{q^2=0}\ .
\end{equation}
Again, applying the gap equations it can be straightforwardly shown that this expression vanishes, proving that the charged technipion is massless.

\begin{figure}[!t]
\includegraphics[width=6.0in]{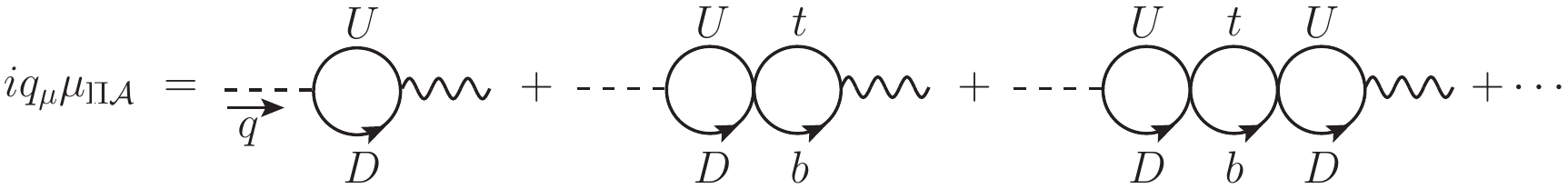}
\caption{Diagrams contributing to the mixing between charged technipion and axial current, at leading order in $N$ and $N_c$. At zero momentum this gives $F_\Pi$ times the square root of the charged-technipion wave-function renormalization.}
\label{Fig:GF}
\end{figure}
Finally, in order to match this chiral-techniquark model to the electroweak theory, we need to impose that the charged-technipion decay constant is equal to $F_\Pi\equiv 246$ GeV. The decay constant can be extracted from the mixing term between the charged technipion and the charged component of the axial current. This is given by the sum of diagrams shown in Fig.~\ref{Fig:GF}, and leads to the expression
\begin{equation}
\mu_{\Pi {\cal A}} = \frac{4\, N\, M}{v}\left[
\mu^{\Lambda_{\rm TC}}(M_U,M)\left(1+\frac{N\, N_c\, G^2\, \chi_{UD}\, \zeta_{tb}}{1-4\, N\, N_c\, G^2\, \zeta_{UD}\, \zeta_{tb}}\right)
+\frac{\mu^{M_{\rm ETC}}(M_t,0)}{2}\frac{N_c\, G\, \chi_{UD}}{1-4\, N\, N_c\, G^2\, \zeta_{UD}\, \zeta_{tb}}\right]\ .
\end{equation}
The function $\mu^\Lambda(M_1,M_2)$ is defined in Eq. (\ref{Eq:mu}).
Using the gap equations, the expression for $\mu_{\Pi {\cal A}}$ can be brought into the compact form
\begin{equation}
\mu_{\Pi {\cal A}}  = 2\left[N\, (M_U+M)\, \mu^{\Lambda_{\rm TC}}(M_U,M)+N_c\, M_t\, \mu^{M_{\rm ETC}}(M_t,0)\right]\ .
\end{equation}
To obtain the decay constant we must divide the latter by the square root of the charged-technipion wave-function renormalization constant:
\begin{equation}
F_\Pi = \displaystyle{\lim_{q^2\to 0}}\displaystyle{\frac{\mu_{\Pi {\cal A}}(q^2)}{\sqrt{\Sigma_{\Pi^-\Pi^+}^\prime(q^2)}}}\ .
\label{Eq:FP}
\end{equation}
It is not difficult to show that, in the limit $G\to 0$, this becomes the Pagels-Stokar relation:
\begin{equation}
F_\Pi^2\, \substack{\ \\ \  \\ {\displaystyle =}\\ G\to 0}\, 
 4\, N\, M^2\, I_2^{\Lambda_{\rm TC}}(M)\ .
\end{equation}
Once $F_\Pi$ is fixed, this implies that small variations of $M$ lead to exponentially large variations of ${\Lambda_{\rm TC}}$, as the ${\Lambda_{\rm TC}}$ dependence in $I_2^{\Lambda_{\rm TC}}(M)$ is essentially logarithmic. Therefore, requiring ${\Lambda_{\rm TC}}$  to be above 2-3 TeV and well below 10 TeV, as expected if ${\Lambda_{\rm TC}}$ is of the order of the heavy technihadron masses, forces $M$ to be within a very narrow interval.

\section{The dynamical mass of the scalar singlet}\label{Sec:scalarmass}
\begin{figure}[!t]
\includegraphics[width=6.0in]{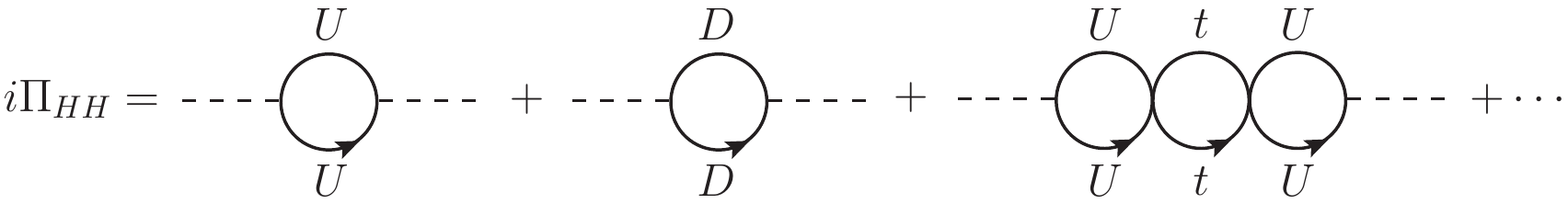}
\caption{Diagrams contributing to the TC-Higgs self-energy at leading order in $N$ and $N_c$.}
\label{Fig:TechniHiggs}
\end{figure}
Next, we will consider the subject of our main interest in this paper, which is the mass of the scalar singlet. Because of the interaction with the top quark, the {\em physical} mass $M_H$ of the scalar singlet may be very different from its {\em dynamical} mass $M_{H0}$: the latter is defined as the mass solely due to the strong TC dynamics, and is obtained by setting $G=0$. Taking also the iteraction with the top quark into account, the scalar-singlet self-energy is given by the diagrams of Fig.~\ref{Fig:TechniHiggs}. Summing the series gives
\begin{equation}
\Sigma_{HH} = -m^2+\frac{N\, y^2\, M^2}{v^2}\left(\xi_D + \xi_U\right)
+\frac{N\, y^2\, M^2}{v^2}\frac{N\, N_c\, G^2\, \xi_U^2\, \xi_t}{1-N\, N_c\, G^2\, \xi_U\, \xi_t}\ ,
\label{higgsself}
\end{equation}
where
\begin{equation}
i\ \xi_X\equiv -\int\frac{d^4k}{(2\pi)^4} {\rm Tr}\  \frac{\slashed{k}+M_X}{k^2-M_X^2} \frac{\slashed{k}+\slashed{q}+M_X}{(k+q)^2-M_X^2}\ .
\end{equation}
Evaluating the integral gives
\begin{eqnarray}
&& \xi_X = 2\left(q^2-4M_X^2\right) I_2^{\Lambda_{\rm TC}}(M_X)+4 I_1^{\Lambda_{\rm TC}}(M_X)\ ,\ \ X=U,D \\
&& \xi_t = 2\left(q^2-4M_t^2\right) I_2^{M_{\rm ETC}}(M_t) + 4 I_1^{M_{\rm ETC}}(M_t)\ .
\end{eqnarray}
We eliminate $m^2$ by requiring the dynamical mass to be $M_{H0}$, i.e. we set $\Sigma_{HH}$ of Eq. (\ref{higgsself}) equal to zero at $q^2=M_{H0}$ at $G=0$. This leads to the expression
\begin{equation}
\Sigma_{HH}=\frac{N y^2 M^2}{v^2} \left(4 I_2^{\Lambda_{TC} } (M) \left(q^2-M_{H0}^2\right)-\xi_D+\frac{\xi_U}{1-N N_c G^2 \xi_U \xi_t}\right)\ .
\end{equation}
We would like to test whether the scalar singlet $H$ can be interpreted as the recently observed 125 GeV resonance. Therefore, we set $\Sigma_{HH}=0$ at $q^2=M_H^2=(125\ {\rm GeV})^2$, and solve for $M_{H0}$. We can then compare the latter with independent estimates for the dynamical mass, such as those obtained from scaling up the QCD spectrum.

\begin{figure}[!t]
\includegraphics[width=3.0in]{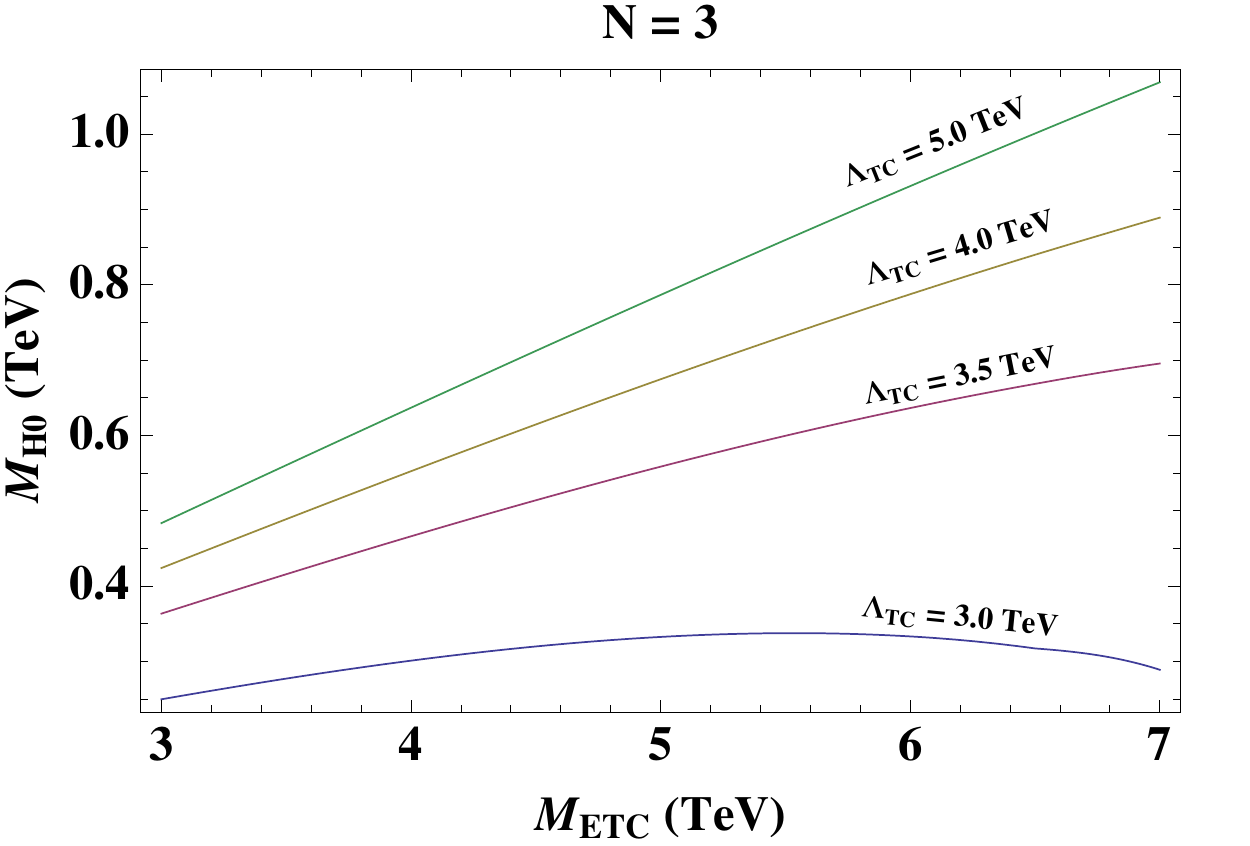}
\includegraphics[width=3.0in]{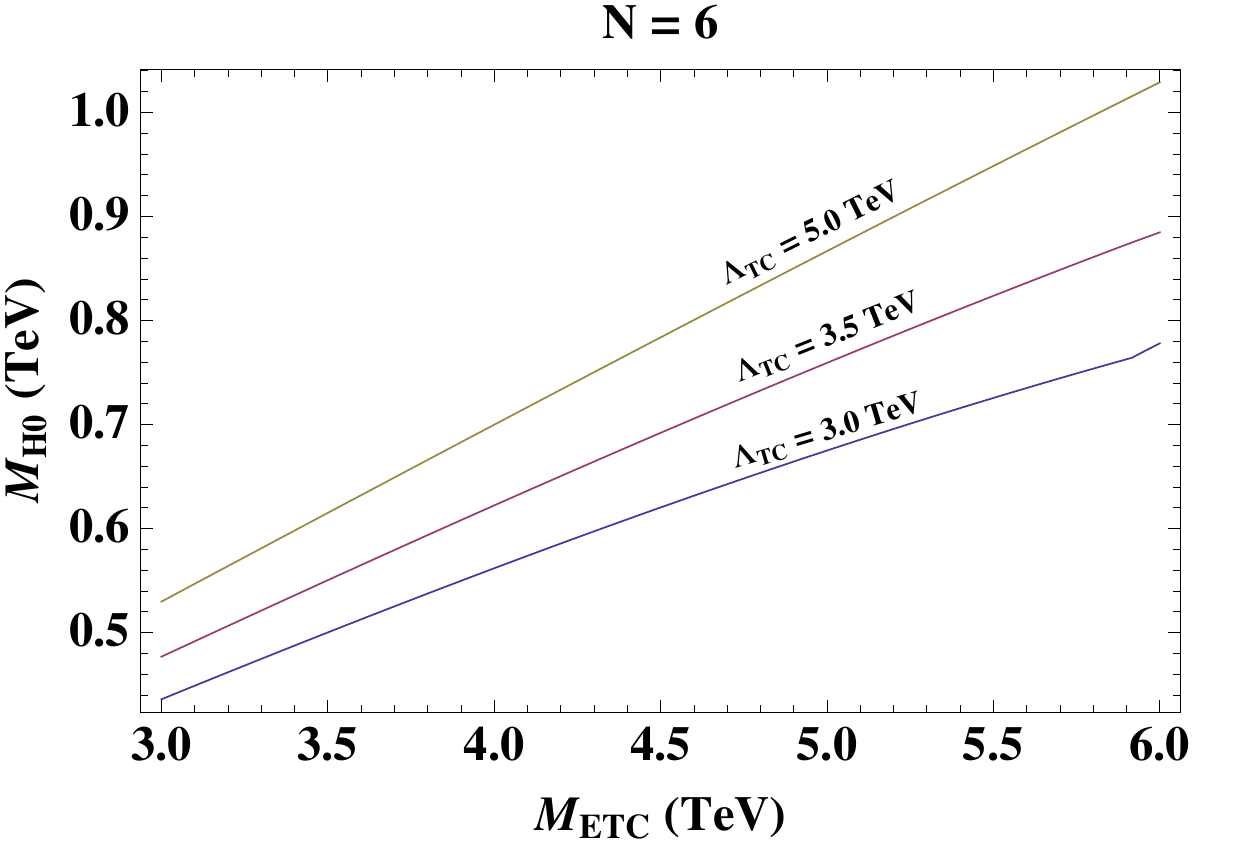}
\caption{The dynamical mass of the scalar singlet, $M_{H0}$, which is required to give a 125 GeV scalar, after the radiative corrections from the top quark are included. For large enough TC scale $\Lambda_{\rm TC}$, $M_{H0}$ grows with the ETC mass $M_{\rm ETC}$, as shown by  (\ref{Eq:Approx}). The left plot is for $N=3$, whereas the right plot is for $N=6$, where $N$ is the dimension of the techniquark representation under the TC gauge group.}
\label{Fig:MH}
\end{figure}

Let us first take the limit $M^2\ll \Lambda_{\rm TC}^2\ll M_{\rm ETC}^2$. Solving for $M_{H0}$ gives, at leading logartithmic order,
\begin{equation}
M_{H0}^2\simeq 
\frac{1}{\log \Lambda_{\rm TC}^2/M^2}\frac{
\displaystyle {\frac{N_c}{N}\frac{M_t^2}{M_U^2}}}
{1-
\displaystyle{\frac{N_c}{N}\frac{M_t^2}{M_U^2}\frac{M_{\rm ETC}^2}{\Lambda_{\rm TC}^2}}}M_{\rm ETC}^2
\,.
\label{Eq:Approx}
\end{equation}
This shows that the dynamical mass required to give a 125 GeV physical mass grows with $M_{\rm ETC}$. Ignoring $M_H^2$ gives a linear growth, as shown in Fig.~\ref{Fig:MH} (left) for $N=3$, and (right) for $N=6$. Note that $M$ and $M_{H0}$ are expected to scale like $1/\sqrt{N}$. We see that $M_{H0}$ may indeed be much larger than 125 GeV.

\section{Results and constraints}\label{Sec:ST}

The results of our analysis can be related to the underlying gauge dynamics of some simple technicolor models. Setting $N=3$ would correspond to the QCD like dynamics of SU(3) gauge theory with two fermions in the fundamental representation of the gauge group. On the other hand $N=6$ would correspond to the dynamics of SU(3) gauge theory with two fermions in the sextet representation. The $N=6$ case has been introduced \cite{Sannino:2004qp} as one of the minimal models with walking dynamics. Its phenomenological viability has been studied e.g. in \cite{Belyaev:2008yj,Dietrich:2005jn,Dietrich:2005wk,Alanne:2013dra} and its nonperturbative properties have been also recently studied on the lattice \cite{DeGrand:2013uha,Fodor:2012ty}

\begin{figure}[h]
\includegraphics[width=3.0in]{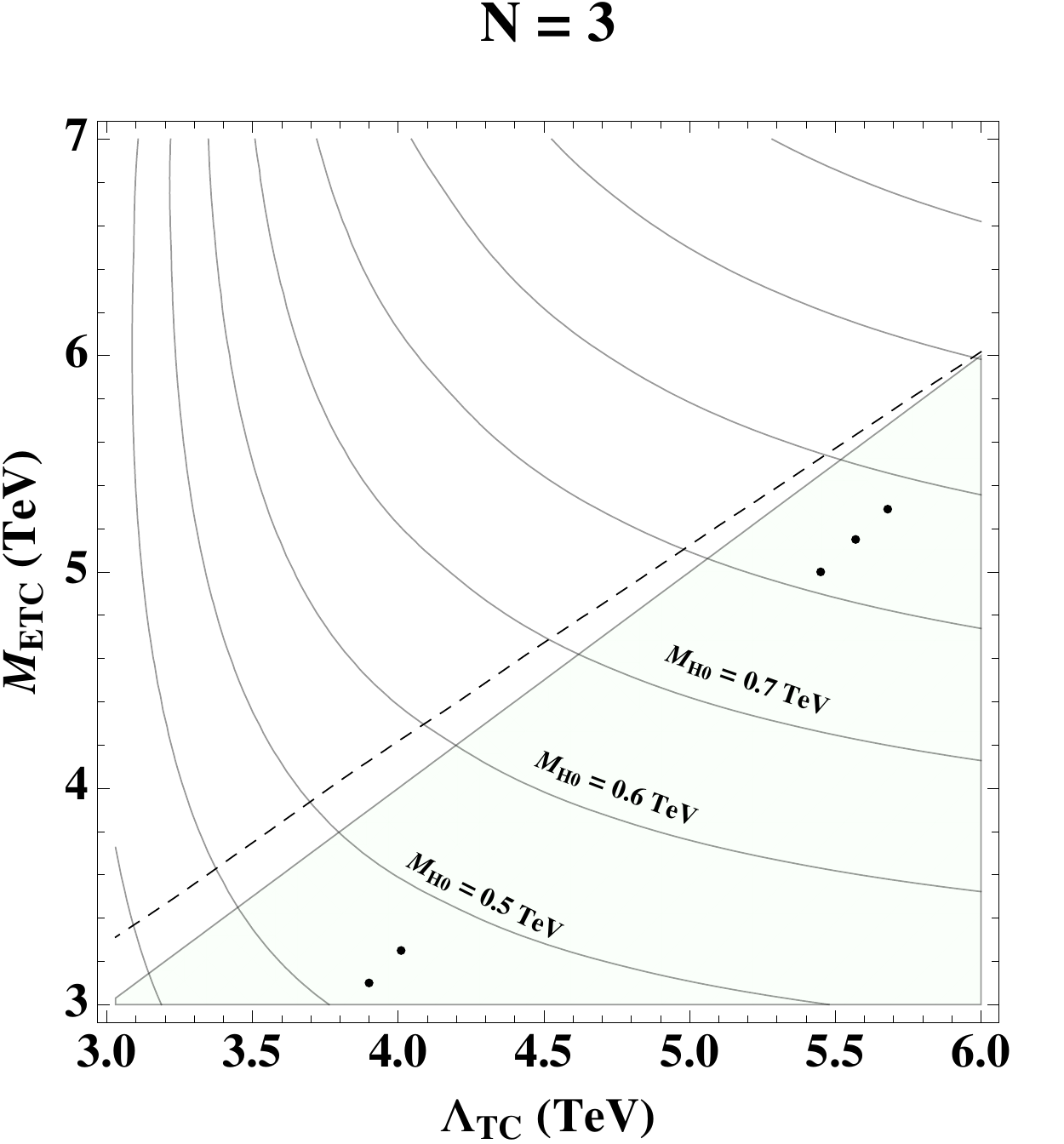}
\includegraphics[width=3.0in]{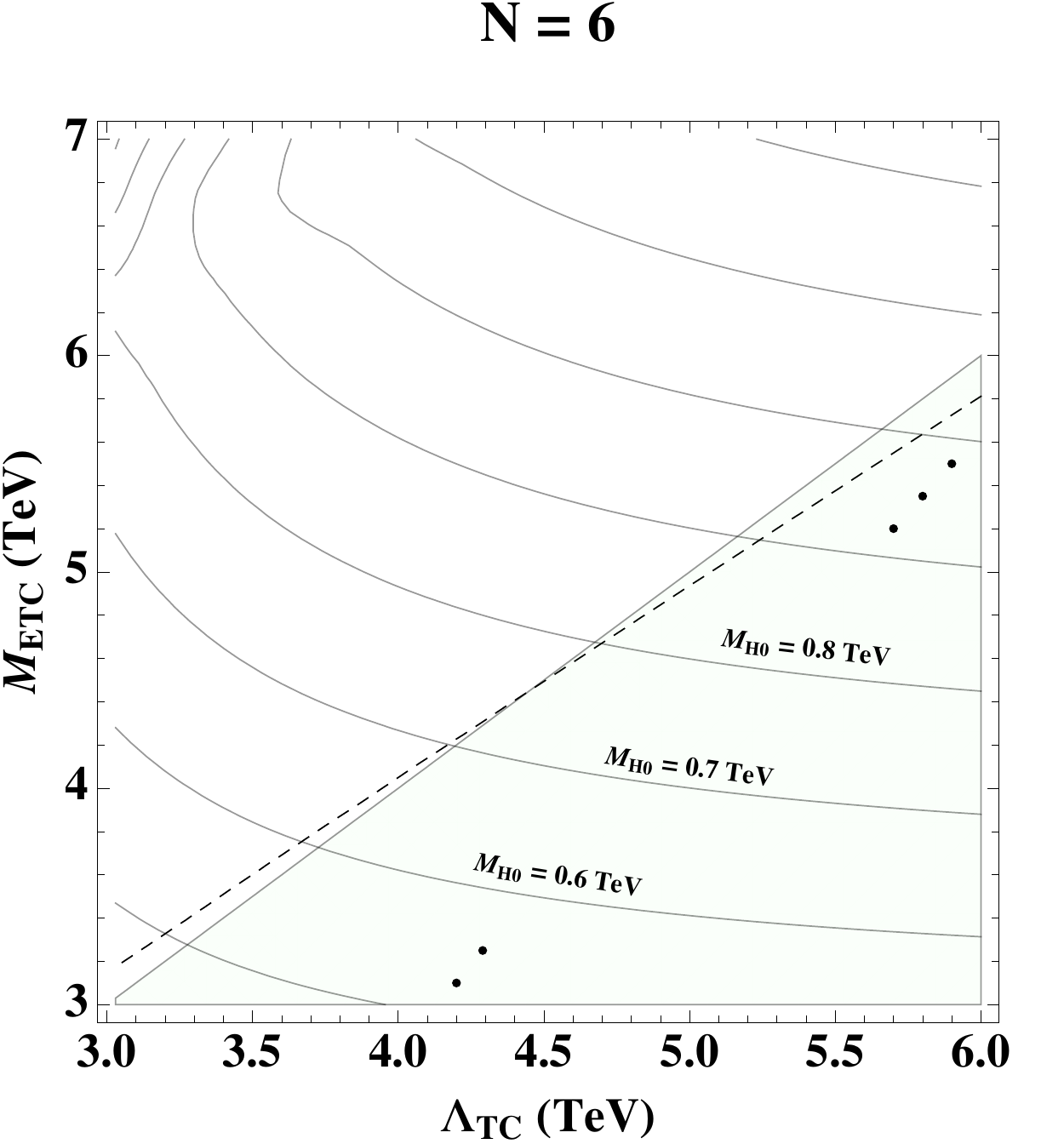}
\caption{The dynamical scalar mass $M_{H0}$ is shown here as contour lines in the  $(\Lambda_{\rm TC},M_{\rm ETC})$ parameter space. Inside the triangualr region in the lower-right corner of the plot, the TC cutoff $\Lambda_{\rm TC}$ is larger than $M_{\rm ETC}$, and our parametrisation of the ETC sector in terms of four-fermion operators breaks down. In the upper-left regions, above the dashed line, the $T$ parameter evaluated using the first term of (\ref{Eq:T}) is greater than its 95\% C.L. upper bound, $T_{\rm up}=0.23$. The left plot is for $N=3$, whereas the right plot is for $N=6$.}
\label{Fig:Viable}
\end{figure}

With these concret models in mind, in Fig.~\ref{Fig:MH} we have shown the dynamical mass of the isosinglet scalar resonance which is required to give a 125 GeV scalar after radiative corrections from the top quark. The values obtained are within several hundreds of GeVs and below $\sim$ 1 TeV. This is a range which can be reasonably expected from a TC theory, especially in the case of non-QCD-like dynamics. The mass scale $\Lambda_{\rm TC}$ is determined by the underlying  strong dynamics, and is ignored in this model computation. The ETC mass scale is a free parameter, and  Fig.~\ref{Fig:MH} shows that larger values of $M_{\rm ETC}$ lead to larger radiative corrections to the scalar mass. However $M_{\rm ETC}$ cannot be arbitrarily large: the first of (\ref{eq:Gap}) shows that the isospin splitting $\delta$ grows with $M_{\rm ETC}$, which should therefore be constrained by the $T$ parameter. In a TC theory with one weak technidoublet the $S$ and $T$ parameters, to leading order in a a large-$N$ expansion, can be written as
\begin{eqnarray}
S &=& \frac{N}{6\pi} + \Delta S_{\rm ETC} + \Delta S_{\rm vectors} \ , \\
T &=& \frac{N}{16\pi\, s_W^2\, c_W^2\, M_Z^2}\, \left[M_U^2+M_D^2-\frac{2\, M_U^2\, M_D^2}{M_U^2-M_D^2}\, \log\frac{M_U^2}{M_D^2}\right] + \Delta T_{\rm ETC} + \Delta T_{\rm vectors} \ , \label{Eq:T}
\end{eqnarray}

where, in the equation for $T$, $M_U$ and $M_D$ are defined to satisfy the gap equations (\ref{eq:Gap}). The first terms in the above equations are the usual one-loop contributions from heavy techniquarks, \cite{Peskin:1990zt,Peskin:1991sw}. The terms $\Delta S_{\rm ETC}$ and $\Delta T_{\rm ETC}$ are corrections due to ETC operators other than (\ref{eq:ETC}). As argued above, these are expected to be of comparable magnitude, and do certainly contribute to low-energy observables (for instance, a four-techniquark operator contributing more to $M_D$ than $M_U$ would reduce the isospin mass splitting introduced by  (\ref{eq:ETC})). Finally, the terms $\Delta S_{\rm vectors}$ and $\Delta T_{\rm vectors}$ are the contributions from the spin-one resonances, which are also expected to be relevant. Also, sub-leading contributions in $1/N$ may be important when $N$ is not too large. It should be noted that additional ETC operators and sub-leading contributions are also expected to affect $M_{H0}$. 
It is therefore not possible to impose 
strict constraints from oblique corrections 
solely based on the Lagrangian $L_{\rm TC}+L_{\rm ETC}$ of (\ref{eq:TC}) and (\ref{eq:ETC}). Nonetheless, it is useful to explore the $(\Lambda_{\rm TC},M_{\rm ETC})$ parameter space. In Fig.~\ref{Fig:Viable} (left) we consider the case $N=3$. Inside the lower-right 
triangular region the TC cutoff $\Lambda_{\rm TC}$ is larger than $M_{\rm ETC}$, and our parametrisation of the ETC sector in terms of four-fermion operators breaks down. This means that in this region we should employ the full ETC theory to evaluate the correction to the mass of the scalar singlet, and the results are necessarily model-dependent. We may take the values of $M_{H0}$ in this region as an estimate for the correct $M_{H0}$. In the upper-left region, above the dashed line, the $T$ parameter evaluated using the first term of (\ref{Eq:T}) is greater than the experimental 95\% C.L. upper bound, $T_{\rm up}=0.23$ derived from the results of the gfitter group \cite{Baak:2012kk}. Finally, the contours correspond to fixed values of $M_{H0}$ as shown by the contour labels in the figure. In Fig.~\ref{Fig:Viable} (right) we show the case $N=6$. We see the that additional contributions to the $T$ parameter should be negative, or else the ETC scale associated to the top mass would be forced to be smaller than a few TeVs.  

\section{Conclusions and outlook}
\label{Sec:conclusions}

It is a novel observation that the dynamical mass of the isosinglet scalar in the spectrum of a strongy coupled gauge theory acquires nontrivial contributions from interactions external to the strong dynamics. A familiar example of such a contribution is the one, due to the electromagnetic interaction, which generates the mass splitting between charged and neutral pions in QCD. In technicolor theories such external interactions are provided by the electroweak sector itself and by any extension aimed at the generation of the masses for the SM matter fields. The profound consequence of this observation, then, is that the lightness of the Higgs boson observed at the LHC experiments can be well compatible with the composite nature of the scalar boson.

In this paper we have carried out a quantitative analysis of this issue within a model of strong dynamics where the technicolor sector is described by a chiral-meson model and the extended technicolor interactions are modelled by an effective four-fermion interaction between the technifermions and the top quark. Restricting the interaction terms to a single four-fermion operator is a simple but sufficient approximation which serves to illustrate our point: we find that 125 GeV scalar can easily result from underlying strong dynamics corresponding to gauge dynamics of SU(3) gauge theory with two Dirac fermions in the sextet representation. 

Our results should be applicable to a wide variety of models utilizing strong dynamics and four-fermion couplings to explain both the electroweak symmetry breaking and the large mass splitting of top and bottom quarks via a dynamical mechanism \cite{Miransky:1988xi,Miransky:1989ds,Bardeen:1989ds,Dobrescu:1997nm,Chivukula:1998wd,He:2001fz,Fukano:2012qx,Fukano:2013kia}. Furthermore, our results together with the first principle determinations of the dynamical mass of the scalar from the lattice \cite{Fodor:2014pqa} will provide constraints on the dynamical models.

The results we have obtained in this paper could be refined by considering a more detailed set of four-fermion operators. Then one could also describe the generation of fermion mass patterns and attempt a more precise determination of the oblique corrections, i.e. $S$ and $T$ parameters. The effect of additional four-fermion interactions on the scalar mass are expected to be qualitatively similar to the case we have considered here. 
\section*{Acknowledgements}
This work was financially supported by the Academy of Finland project 267842.

\appendix
\section{Integrals}
The standard integrals used in our computations are
\begin{eqnarray}
I_1^{\Lambda}(M) &\equiv&   i\int\frac{d^4 k}{(2\pi)^4}\frac{1}{k^2-M^2}=\frac{1}{16\pi^2}\left({\Lambda}^2-M^2\log\frac{{\Lambda}^2+M^2}{M^2}\right)\ ,  \label{Eq:I1} \\
&&\ \nonumber \\
I_2^{\Lambda}(M) &\equiv &  -i\int\frac{d^4 k}{(2\pi)^4}\frac{1}{(k^2-M^2)^2} = \frac{1}{16\pi^2}\left(\log\frac{{\Lambda}^2+M^2}{M^2}-\frac{{\Lambda}^2}{{\Lambda}^2+M^2}\right)\ ,  \label{Eq:I2} \\
&&\ \nonumber \\
I_2^{\Lambda}(M_X,M_Y) &\equiv & -i \int_0^1 dx \int\frac{d^4 k}{(2\pi)^4}\frac{1}{\left(k^2-x\ M_X^2-(1-x)M_Y^2\right)^2} \nonumber \\
&=&\frac{1}{16\pi^2(M_X^2-M_Y^2)} \left(M_X^2\log\frac{{\Lambda}^2+M_X^2}{M_X^2}-M_Y^2\log\frac{{\Lambda}^2+M_Y^2}{M_Y^2}\right) \ , \label{Eq:I2b} \\
&&\ \nonumber \\
\mu^{\Lambda}(M_X,M_Y) &\equiv& -i\int_0^1 dx \int\frac{d^4 k}{(2\pi)^4} \frac{x\ M_X+(1-x)M_Y}{\left(k^2-x\ M_X^2-(1-x)M_Y^2\right)^2} \nonumber \\
&=& \frac{1}{32\pi^2(M_X^2-M_Y^2)(M_X+M_Y)}\Bigg({\Lambda}^4\log\frac{{\Lambda}^2+M_X^2}{{\Lambda}^2+M_Y^2}
-(M_X^2-M_Y^2){\Lambda}^2  \nonumber \\
&+&M_X^3(M_X+2M_Y)\log\frac{{\Lambda}^2+M_X^2}{M_X^2}
-M_Y^3(2M_X+M_Y)\log\frac{{\Lambda}^2+M_Y^2}{M_Y^2} \Bigg)\ . \label{Eq:mu}
\end{eqnarray}
Note that the last integral is only logarithmically divergent, for ${\Lambda}\to\infty$.
\bibliographystyle{aipnum4-1}

\bibliography{HiggsETC.bib}

\end{document}